\documentclass[12pt,twoside]{article}
\usepackage[mathscr]{eucal}
\usepackage{amsmath,amsfonts,amssymb,amsthm,mathabx,empheq}
\usepackage{times}
\usepackage{pdfsync}
\usepackage{cite}
\usepackage{url}
\usepackage{hyperref}
\usepackage{tensor}
\usepackage{color}
\usepackage{multicol}
\usepackage{bbold}

\advance\textheight\topskip
\DeclareMathOperator\arctanh{arctanh}

\allowdisplaybreaks[1]

\newcommand\td{\text{d}}

\newcommand{\p}{\partial}

\newcommand{\be}{\begin{equation}}
\newcommand{\ee}{\end{equation}}
\newcommand{\bea}{\begin{eqnarray}}
\newcommand{\eea}{\end{eqnarray}}
\def\nn{\nonumber}
\def\bz{\bar z}

\def\half{\frac12}

\def\bY{\bar Y}

\def\bm{\bar{m}}

\def\bL{\bar{L}}

\def\bomega{\bar\omega}

\def\bL{\bar{L}}
\def\bb{\bar{b}}
\def\ba{\bar{a}}

\newcommand*\xbar[1]{%
  \hbox{%
    \vbox{%
      \hrule height 0.5pt 
      \kern0.3ex
      \hbox{%
        \kern-0.0em
        \ensuremath{#1}%
        \kern-0.0em
      }%
    }%
  }%
}

\DeclareFontFamily{OT1}{rsfs}{} \DeclareFontShape{OT1}{rsfs}{m}{n}{
<-7> rsfs5 <7-10> rsfs7 <10-> rsfs10}{}
\DeclareMathAlphabet{\mycal}{OT1}{rsfs}{m}{n}

\begin{document}
\title{Notes on self-dual gravity}

\author{Pujian Mao and Weicheng Zhao}

\date{}

\def\mytitle{Notes on self-dual gravity}

\addtolength{\headsep}{4pt}

\begin{centering}

  \vspace{1cm}

  \textbf{\Large{\mytitle}}

  \vspace{1.5cm}

  {\large Pujian Mao and Weicheng Zhao}

\vspace{.5cm}

\vspace{.5cm}
\begin{minipage}{.9\textwidth}\small \it  \begin{center}
     Center for Joint Quantum Studies and Department of Physics,\\
     School of Science, Tianjin University, 135 Yaguan Road, Tianjin 300350, China
 \end{center}
\end{minipage}

\end{centering}


\vspace{1cm}

\begin{center}
\begin{minipage}{.9\textwidth}
\textsc{Abstract}. In this paper, we study  self-dual gravity in the Newman-Penrose formalism. We specify the self-dual solution space from the Newman-Unti solutions. We show that the asymptotic symmetries of the self-dual gravity are still the (extended) BMS symmetries. We transform the self-dual Taub-NUT solution into the Newman-Unti gauge in analytical form.
 \end{minipage}
\end{center}
\thispagestyle{empty}


\section{Introduction}

Physics at the asymptotic boundary of spacetime has obtained renewed interest in recent years since the seminal work of Barnich and Troessaert \cite{Barnich:2009se,Barnich:2010eb,Barnich:2010ojg} where the asymptotic symmetry algebra at null infinity in 4 dimensions is shown to consist of the semidirect sum of supertranslations with superrotations, the (local) conformal symmetry of the celestial sphere which extends the original BMS symmetry \cite{Bondi:1962px,Sachs:1962wk}. Considering  asymptotic symmetries as symmetries of S-Matrix in quantum gravity, the Ward identities of the supertranslations and superrotations recover the leading and subleading soft graviton theorems respectively \cite{Strominger:2013jfa,He:2014laa,Kapec:2014opa,Campiglia:2014yka}. Alternatively, one can think that any soft graviton theorem is indicating the existence of some symmetries of quantum theory of gravity. If one considers only the tree level plus-helicity soft particles, the entire tower of soft graviton theorems implies a chiral Kac-Moody symmetry of the wedge algebra of $w_{1+\infty}$  \cite{Guevara:2021abz,Strominger:2021lvk}. Those symmetries even persist at all loop level  \cite{Ball:2021tmb} when restricted to the self-dual sector, i.e., quantum self-dual gravity which is a theory of gravity with half of the components of the Weyl tensor vanishing. Although the self-dual structure in four dimensions excludes real solutions in Lorentzian signature, it provides a good toy model to demonstrate some quantum aspects of gravity from different perspectives such as integrablility, string theories and scattering amplitudes (see, e.g.,
\cite{Ward:1985gz,Mason:1991rf,Ooguri:1990ww,Ooguri:1991fp,Kiritsis:1993pb,Bianchi:1994gi,Berkovits:1994vy,Berkovits:1994ym,Duff:1994an,Berkovits:1995ab,Ooguri:1995cp,Bern:1998xc,Monteiro:2011pc,Monteiro:2014cda,Krasnov:2016emc,Berman:2018hwd} and references therein). However a detailed investigation of asymptotic symmetries of the self-dual gravity in the standard asymptotic analysis is still missing in the literature. Considering the importance of asymptotic symmetries in the recent developments introduced above, the aim of the present work is to explore this more in detail.

In \cite{Campiglia:2021srh}, asymptotic symmetries of self-dual gravity were derived in a light-cone gauge in the context of the classical double copy and only a subset of the BMS symmetry is preserved by the self-dual condition. While this result is somewhat surprising if one considers electromagnetism for a simpler analogue of gravitational effects. The self-dual condition is on the field strength which can not bring any constraint on a gauge transformation. In this paper, we study the asymptotic symmetries of self-dual gravity in the Newman-Penrose (NP) formalism. Adopting the widely used Newman-Unti (NU) gauge \cite{Newman:1962cia}, we recover the full (extended) BMS symmetry that preserves the self-dual condition at null infinity.

Another line of research about asymptotic symmetries is their applications in deriving the Bekenstein-Hawking
entropy for black holes \cite{Brown:1986nw,Strominger:1997eq,Guica:2008mu}.
The study of asymptotic symmetries normally relies on some particular coordinates system with desired gauge conditions, e.g., the Bondi gauge \cite{Bondi:1962px} or the NU gauge adopted in the present work. So it is very important to have the black hole solutions in the coordinates system where the asymptotic symmetries are studied. Another purpose of this paper is to give a generic treatment to transform solutions into the NU gauge and we demonstrate the derivation in the case of the self-dual Taub-NUT solution \cite{Taub:1950ez,Newman:1963yy}.

This paper is organized as follows. In the next section, we specify the self-dual conditions on the solutions of the NP equations. In section 3, we show that the  self-dual conditions do not bring any constraint on the asymptotic symmetries. The asymptotic symmetries of the self-dual gravity are still the full (extended) BMS symmetries. In section 4, we introduce a generic treatment transforming solutions into the NU gauge and write the self-dual Taub-NUT solutions precisely in the NU gauge. The last section is devoted to the conclusion and discussion for future directions. There are three appendices which provide useful information for the computations in the main text.

\section{The self-dual sector in NP formalism}

The Newman-Penrose formalism \cite{Newman:1961qr} is a special tetrad formalism with four null basis vectors $e_1=l=e^2,\;e_2=n=e^1,\;e_3=m=-e^4,\;e_4=\bar{m}=-e^3$. In Lorentzian signature, the basis vectors $m$ and $\bm$ are complex and conjugates of each other. The basis vectors are orthogonalized and normalized as
\be\label{tetradcondition}
l\cdot m=l\cdot\bm=n\cdot m=n\cdot\bm=0,\quad l\cdot n=1,\quad m\cdot\bm=-1.
\ee
The spacetime metric is constructed from the tetrad as
\be
g_{\mu\nu}=n_\mu l_\nu + l_\mu n_\nu - m_\mu {\bm}_\nu - m_\nu \bm_\mu.
\ee
For the NP variables and equations, we would refer to \cite{Chandrasekhar} for the notations.

Newman and Unti \cite{Newman:1962cia} derived a generic solution of the NP system with certain gauge and boundary conditions. First, by local Lorentz transformations, they set
\begin{align}\label{connetiongauge}
\pi=\kappa=\epsilon=0,\,\,\;\;\rho=\bar\rho,\;\;\,\,\tau=\bar\alpha+\beta.
\end{align}
Such gauge choices mean that $l$ is tangent to a null geodesic with affine parameter, the rest basis vectors are parallelly transported along $l$, and $l$ is the gradient of a scalar field. In this case, one can choose the scalar field and the affine parameter from $l$ as coordinates $x^1=u$ and $x^2=r$. For the rest two angular coordinates, the complex stereographic coordinates $x^A=(z,\bz)$ were applied. They are related to the usual angular variables $(\theta,\phi)$ by $z=\cot\frac\theta2 e^{i\phi}$. In these coordinates, the tetrad and the co-tetrad satisfying conditions in \eqref{tetradcondition} must have the forms
\be\label{tetradgauge}
\begin{split}
&n=\frac{\p}{\p u} + U \frac{\p}{\p r} + X^A \frac{\p}{\p x^A},\\
&l=\frac{\p}{\p r},\\
&m=\omega\frac{\p}{\p r} + L^A \frac{\p}{\p x^A},\\
&\bm=\xbar\omega\frac{\p}{\p r} + \bL^A \frac{\p}{\p x^A},
\end{split}
\ee
and
\be\label{co-tetradgauge}
\begin{split}
&n=\left[-U-X^A(\xbar\omega L_A+\omega \bar L_A)\right]\td u + \td r + (\omega\bar L_A+\xbar\omega L_A) \td x^A,\\
&l=\td u,\\
&m=-X^A L_A \td u + L_A \td x^A,\\
&\bm=-X^A \bL_A \td u + \bL_A \td x^A,
\end{split}
\ee
where $L_AL^A=0,\;L_A\bar L^A=-1$. The asymptotic behaviors of the NP variables imposed in \cite{Newman:1962cia} to derive the solution are\footnote{Originally, the boundary conditions in \cite{Newman:1962cia} are $\rho=-r^{-1} + O(r^{-2}),\tau,\xbar\tau=O(r^{-1}),X^A=O(1)$. However those conditions can be made stronger by pure gauge transformations. So we think it is reasonable to impose them in the beginning. The rest boundary conditions for the NP variables in \cite{Newman:1962cia} are not really needed for deriving the solution. They could be the consequence of the NP equations.}
\be\label{boundarycondition}
\begin{split}
&\rho=-r^{-1} + O(r^{-3}),\quad \sigma,\xbar\sigma=O(r^{-2}),\quad \tau,\xbar\tau=O(r^{-2}),\\
&L^{\bz},\bL^z=O(r^{-1}),\quad \bL^{\bz},L^{z}=O(r^{-2}),\quad  X^A=O(r^{-1}),\quad \Psi_0=O(r^{-5}).
\end{split}
\ee
Those conditions determine an asymptotically flat spacetime in Lorentzian signature.

The NP formalism is originally designed to investigate systems with gravitational radiation. Hence a pair of complex null bases is needed to keep the spacetime in Lorentzian signature. Nevertheless it is more natural to adopt the $(2,2)$ Kleinian signature in the NP formalism. In the Kleinian signature, one can just work with four independent real basis vectors, i.e., $m$ and $\bm$ are now real and independent. Correspondingly, $z$ and $\bz$ are independent real coordinates and all NP variables are independent of the ones with an overhead bar. Analytic continuation from Lorentzian to Kleinian signature has seen a surge in interest recently from both quantum aspect (see, e.g., \cite{Elvang:2013cua,Henn:2014yza}) and classical aspect \cite{Atanasov:2021oyu,Crawley:2021auj}. A more relevant fact to our work is that it admits self-dual solutions in Kleinian signature.

We will deal with vacuum Einstein gravity without cosmological constant in this work. So the Riemann tensor and the Weyl tensor are the same. The components of the Weyl tensor are either self-dual under Hodge's star operation
\be
^\ast C_{abcd}=\frac12 \epsilon_{abef}{C^{ef}}_{cd},
\ee
or anti-self-dual under Hodge's star operation
\be
^\ast C_{abcd}=-\frac12 \epsilon_{abef}{C^{ef}}_{cd} .
\ee
It is easy to verify in NP variables that $^\ast\Psi_i=\Psi_i$ and $^\ast\xbar\Psi_i=-\xbar\Psi_i$. Hence the self-dual sector in the NP formalism is just to impose that $\xbar\Psi_i=0$. Clearly it is very convenient to study the self-dual gravity in the NP formalism where the Weyl tensor is included in the equations of motion. One can directly check the self-dual conditions in the solutions.

The self-dual solutions can be obtained from the analytic continuation of the Newman-Unti solutions by simply setting $\xbar\Psi_i=0$. The self-dual solutions in series expansion is given in Appendix \ref{solution}. In particular, we set $L^{\bz},\bL^z=\frac1r + O(r^{-2})$, i.e., the null infinity as the product of a null interval with a plane or a torus after compactification in $z$ and $\bz$ directions. It is a more convenient choice to demonstrate the self-dual conditions in solution space. The generic case of $L^{\bz},\bL^z= O(r^{-1})$ can be obtained by a Weyl transformation at the infinity, see, e.g., the discussions in \cite{Barnich:2016lyg,Compere:2016jwb,Ball:2018prg,Barnich:2021dta}. The precise constraints from the self-dual conditions are that $\xbar\lambda^0$ is only a function of $\bz$. Consequently, $\sigma^0$ is determined as
\be
\sigma^0=u \xbar\lambda^0(\bz)  + \Theta(z,\bz).
\ee
It is worthwhile to point out that the non-vanishing of $\sigma^0$ and $\xbar\lambda^0$ under the self-dual conditions saves the full BMS symmetry as we will present in the next section.

\section{Asymptotic symmetries in self-dual sector}

In 1960s, Bondi and collaborators formulated the Einstein equation as a characteristic initial value problem for axisymmetric isolated systems to understand the gravitational radiation in full Einstein theory \cite{Bondi:1962px}. Surprisingly, they found that the asymptotic symmetry group in this system has
an infinite dimensional subgroup, i.e., the supertranslations which are translations in the time direction involving an arbitrary function of the angular variable. Later on, Sachs extended this system to the generic asymptotically flat case \cite{Sachs:1962wk}. The asymptotic symmetries of Sachs's system are the semidirect sum of supertranslations with the Lorentz symmetries. This is the original version of the BMS symmetry. In the NU gauge, the BMS symmetry was recovered in \cite{Newman:1962cia}. More than ten years ago, Barnich and Toessaert reported that if one removes some conditions from global aspects, the Lorentz part of the BMS symmetry should be replaced by local conformal symmetries of the celestial sphere which is referred to as superrotations. The full symmetry algebra is the semidirect sum of supertranslations with two copies of Virasoro algebra \cite{Barnich:2009se,Barnich:2010eb,Barnich:2010ojg}. The connection of BMS symmetry in the Bondi gauge and in the NU gauge was studied in \cite{Barnich:2011ty}. Other extensions of the BMS symmetry can be found, e.g., in \cite{Campiglia:2014yka,Compere:2018ylh,Campiglia:2016efb,Freidel:2021dfs}. The emergence of the full infinite-dimensional local conformal symmetries indicates that four dimensional quantum gravity should be closely related to two dimensional conformal field theory. Such idea was first realized by connecting 4d scattering amplitudes to 2d correlators  \cite{Pasterski:2016qvg,Pasterski:2017kqt}. The subset of 2d CFT correlators that is connected to 4d amplitudes are normally referred to as celestial amplitudes \cite{Pasterski:2021rjz} which is at the core of the recently established flat holography  \cite{Ball:2019atb,Pasterski:2021raf}.

In the NP formalism, the gauge transformation is a
combination of a diffeomorphism and a local Lorentz transformation. Let ${\xi}^\mu$ and ${{\omega}^{a}}_b=-{\omega_b}^a$
denote parameters for the infinitesimal transformations, they act on the NP variables as
\begin{equation}
\begin{split}
&\delta_{\xi,\omega}{e_a}^\mu ={\xi}^\nu\partial_\nu
    {e_a}^\mu-\p_\nu{\xi}^\mu{e_a}^\nu +{\omega_a}^b{e_b}^\mu,\\
&\delta_{\xi, \omega} \Gamma_{a b c} = {\xi}^\nu \partial_\nu \Gamma_{a b c} - e_c^\mu \p_\mu {\omega}_{a b} + {\omega_a}^{d}\Gamma_{dbc}+ {\omega_b}^{d}\Gamma_{adc} + {\omega_c}^{d}\Gamma_{abd},\\
&\delta_{\xi, \omega} C_{abcd} ={\xi}^\nu \partial_\nu C_{abcd}
+ {{\omega}_a}^f C_{fbcd} + {{\omega}_b}^f C_{afcd}
+ {{\omega}_c}^f C_{abfd} + {{\omega}_d}^f C_{abcf}.
\end{split}
\end{equation}
The parameters of residual gauge transformations that preserve the NU solution space are entirely determined by asking the conditions \eqref{connetiongauge} and
\eqref{tetradgauge} and \eqref{boundarycondition} to be preserved on-shell. This is worked out in detail in \cite{Barnich:2019vzx}. Adapted to our conventions, the asymptotic symmetry parameters are characterized by three arbitrary functions
\begin{equation}
  T(z,\bz),\quad Y^z=Y(z),\quad
  Y^{\bz}= \bY(\bz).
  \label{orig}
\end{equation}
They represent the supertranslations and superrotations respectively. The associated residual gauge transformations are explicitly determined by the above parameters as
\begin{equation}
\begin{split}
&\xi^u =f(u,z,\bz)=T(z,\bz)+\frac12 u (\p_z Y + \p_{\bz} \bY),\\
&\xi^A= Y^A(x^A) - \p_B f \int^{+\infty}_r dr[L^A \bL^B +  \bL^AL^B ], \\
&\xi^r=-\p_u f r + \p_z \p_{\bz} f - \p_A f
\int^{+\infty}_r dr[\omega \bL^A + \bomega L^A + X^A], \\
\end{split}
\end{equation}
and
\begin{equation}
\begin{split}
\omega^{12}=& \p_u f + X^A \p_A f, \quad
\omega^{23}= \bL^A \p_A f,\quad
\omega^{24}= L^A \p_A f, \\
\omega^{13}=& -
 \partial_u \p_z f + \p_A f \int^{+\infty}_r dr[\lambda
L^A + \mu \bL^A], \\
\omega^{14}=& -
\partial_u \p_{\bz} f + \p_A f \int^{+\infty}_r dr[\bar{\lambda}
\bL^A + \bar{\mu} L^A], \\
\omega^{34}=& \frac12 ( \p_{\bz} \bY - \p_z Y ) - \partial_A f
\int^{+\infty}_{r} dr [(\bar{\alpha} - \beta ) \bar{L}^A + (\bar{\beta} - \alpha) L^A].
\end{split}
\end{equation}
To specify the constraint from the self-dual conditions, one just needs to work out the transformation law of the relevant NP variables. In particular, for the $\xbar\Psi_i$ part, their transformations are
\be
\begin{split}
&\delta_s \xbar\Psi_0 = \xi^\mu \p_\mu \xbar \Psi_0 + 2(\omega^{21} + \omega^{43}) \xbar \Psi_0 + 4 \omega^{23}\xbar\Psi_1,\\
& \delta_s \xbar\Psi_1 = \xi^\mu \p_\mu \xbar \Psi_1 + (\omega^{21} + \omega^{43}) \xbar \Psi_1 + 3 \omega^{23}\xbar\Psi_2 + \omega^{14}\xbar\Psi_0,\\
&  \delta_s \xbar\Psi_2 = \xi^\mu \p_\mu \xbar \Psi_2 + 2 \omega^{23}\xbar\Psi_3 + 2\omega^{14}\xbar\Psi_1,\\
&\delta_s \xbar\Psi_3 = \xi^\mu \p_\mu \xbar \Psi_3 - (\omega^{21} + \omega^{43}) \xbar \Psi_3 +  \omega^{23}\xbar\Psi_4 + 3\omega^{14}\xbar\Psi_2,\\
&\delta_s \xbar\Psi_4 = \xi^\mu \p_\mu \xbar \Psi_4 -2 (\omega^{21} + \omega^{43}) \xbar \Psi_4 + 4 \omega^{14}\xbar\Psi_3.\\
\end{split}
\ee
where $s=(T,Y,\bY)$. Since there is no inhomogenous term, the conditions $\xbar\Psi_i=0$ can not bring any constraint on the residual gauge transformations. One can show that the transformation laws of the rest fields are
\begin{equation}\label{transformationlaw}
\begin{split}
&\delta_s \xbar\sigma^0=[Y \p_z +\bY \p_{\bz} + f \p_u + \p_u f - \p_{\bz} \bY + \p_z Y]\xbar\sigma^0 - \p_z^2 f,\\
&\delta_s \lambda^0=[Y \p_z +\bY \p_{\bz} + f \p_u +  2\p_u f
- \p_{\bz} \bY + \p_z Y]\lambda^0 - \frac12 \p_z^3 Y,\\
&\delta_s \Psi^0_0=[Y \p_z +\bY \p_{\bz} + f \p_u + 3\p_u f +\p_{\bz} \bY - \p_z Y]\Psi^0_0 + 4 \Psi^0_1 \p_{\bz} f,\\
&\delta_s \Psi^0_1=[Y \p_z +\bY \p_{\bz} + f \p_u + 3\p_u f +\frac12(\p_{\bz} \bY - \p_z Y)]\Psi^0_1 + 3 \Psi^0_2 \p_{\bz} f,\\
&\delta_s \Psi^0_2=[Y \p_z +\bY \p_{\bz} + f \p_u + 3\p_u f
]\Psi^0_2 + 2 \Psi^0_3 \p_{\bz} f,
\end{split}
\end{equation}
and
\be
\begin{split}
&\delta_s \Theta=[Y \p_z +\bY \p_{\bz}  + 2 \p_{\bz} \bY]\Theta - \p_{\bz}^2 T,\\
&\delta_s \xbar\lambda^0=[\bY \p_{\bz} +  2\p_{\bz} \bY]\xbar\lambda^0 - \frac12 \p_{\bz}^3 \bY.
\end{split}
\ee
It is clear from the transformation law that if $\Theta$ and $\xbar\lambda^0$ are vanishing, the asymptotic symmetries will be changed significantly. One copy of the Virasoro symmetry will be reduced to $SL(2,R)$ and the supertranslations are reduced to two holomorphic functions which are solutions of $\p_{\bz}^2 T=0$. This will precisely recover the symmetries obtained in \cite{Campiglia:2021srh}.

The charges generating the BMS transformations in the self-dual case take the form \cite{Barnich:2019vzx}
\begin{multline}
{\cal Q}_s=-\frac{1}{8\pi G}\int_{\Gamma}f(\Psi^0_2 + \sigma^0
\lambda^0 + \xbar\sigma^0
\xbar\lambda^0) +Y[\half
\p_z(\sigma^0\xbar\sigma^0)  +\xbar\sigma^0\p_z\sigma^0 ] \\ +\bY[ \half
\p_{\bz}(\sigma^0\xbar\sigma^0)  +\sigma^0\p_{\bz}\xbar\sigma^0+ \Psi^0_1],
\end{multline}
where $\Gamma$ can be any two-surface at fix time at the null infinity. Two immediate remarks follow from the expressions of the charges. First, the charges associated with full superrotations are not vanishing. This means that the self-dual condition can not reduce the asymptotic symmetries at the charge level either. Second, it is very important to recover the full BMS symmetry for the self-dual theory from the charge analysis. $\Psi_1^0$ as the angular momentum aspect in the full theory arises only in the charges of the anti-holomorphic part of the superrotations which are reduced to $SL(2,R)$ in the subset BMS symmetry \cite{Campiglia:2021srh}.

\section{Self-dual Taub-NUT solutions in NU gauge}

A special feature of the NU gauge is that it is based on the congruence of null geodesics that is hypersurface orthogonal and the generator of the null geodesics is the gradient of a scalar field. Once a solution admits such null vector in explicit form, the solution can be written in the NU gauge in a straightforward way, see also  \cite{2003CQGra..20.4153F} for relevant discussion for the Kerr case. So the crucial step to having solutions in the NU gauge is to find the null geodesics generator $l$. This can be done systematically in the NP formalism. One can split the task into two steps. Once the solution is written in the NP formalism with four null basis vectors, the first step is to use three classes of tetrad rotations to turn off certain spin coefficients, namely conditions in \eqref{connetiongauge}. Note that there are the other half conditions for the variables with a bar. Such conditions will guarantee that the null basis $l$ generates null geodesics and is the gradient of a scalar field. The second step is to use coordinates transformation to have the basis vectors in the NU gauge \eqref{co-tetradgauge}. Since the spin coefficients are scalars, there are not changed under any coordinates transformation.

According to the transformation property of the three classes of rotations which are listed in Appendix \ref{rotations} in Kleinian signature, the strategy for the first step is as follows. First, one needs to perform a combined third and second classes of rotations to set $\rho=\xbar\rho$ and $\xbar\kappa=\kappa=0$. While only a second class of rotation should be involved to achieve so in some special cases. Then one needs to use a third class of rotation to turn off $\epsilon$ and $\xbar\epsilon$. Next, a first class of rotation will turn off $\pi$ and $\xbar\pi$. At last, another third class of rotation may be involved to set $\tau=\xbar\alpha+\beta$ and $\xbar\tau=\alpha+\xbar\beta$. The last operation will not turn on $\epsilon$ nor $\xbar\epsilon$. Once the spin coefficient conditions are fulfilled, there must be a coordinates system that will have the basis vectors in the form of \eqref{tetradgauge} and \eqref{co-tetradgauge}.

We will test the above processes by transforming the self-dual Taub-NUT solution into the NU gauge. The Taub-NUT solution in Lorentzian signature in the complex stereographic coordinates adapted to the NP conventions is \cite{Griffiths:2009dfa}
\be\label{Taub-NUT-metric}
\td s^2 = f(r)\left(\td t + 2 i N \frac{z\td \bz - \bz \td z}{1+z\bz}\right)^2 - \frac{\td r^2}{f(r)} - (r^2+N^2)\frac{4\td z \td \bz}{(1+z\bz)^2},
\ee
where $f(r)=\frac{r^2 - 2 Mr - N^2}{r^2 + N^2}$. $N$ is the NUT parameter and $M$ is the mass parameter. When setting $N=0$, one recovers the Schwarzschild solution. One can obtain the solution in Kleinian signature simply by defining a real function ${\cal N}=i N$ and considering $z$ and $\bz$ as independent real coordinates. The Taub-NUT solution is of Petrov type D. Hence the only non-vanishing Weyl scalars are $\Psi_2$ and $\xbar\Psi_2$ as shown in Appendix \ref{taubnutnp}. In the Kleinian signature, one has
\be
\Psi_2=-\frac{M + {\cal N}}{(r+{\cal N})^3}, \quad \xbar\Psi_2=-\frac{M - {\cal N}}{(r-{\cal N})^3}.
\ee
The self-dual solution is simply to require ${\cal N}=M$. The line element of the self-dual solution is
\be
\td s^2 = f(r)\left(\td t + 2M \frac{z\td \bz - \bz \td z}{1+z\bz}\right)^2 - \frac{\td r^2}{f(r)} - (r^2- M^2)\frac{4\td z \td \bz}{(1+z\bz)^2}.
\ee
Now one has $f(r)=\frac{r-M}{r+M}$. We define $\tilde u=t-r - 2M \ln(r-M)$ and choose the following four null bases
\be
\begin{split}
&l=\td \tilde u - 2M \frac{\bz \td z}{1+z\bz} +  2M \frac{z\td \bz}{1+z\bz} ,\\
&n= f \left(\frac12\td \tilde u + \frac{\td r}{f} -  M \frac{\bz \td z}{1+z\bz} + M \frac{z\td \bz}{1+z\bz} \right),\\
&m=\sqrt2(r - M)\frac{\td z }{(1+z\bz)},\\
&\bm=\sqrt2(r +M)\frac{\td \bz}{(1+z\bz)}.
\end{split}
\ee
In the vector form, they are
\be
\begin{split}
&l=\p_r,\\
&n= \p_{\tilde u} - \frac{f}{2} \p_r,\\
&m=\frac{\sqrt2 M z}{r+M}\p_{\tilde u} -  \frac{1+z \bz}{\sqrt2 (r+M)}\p_{\bz},\\
&\bm=\frac{\sqrt2 M \bz}{M-r}\p_{\tilde u} -  \frac{1+z \bz}{\sqrt2 (r-M)}\p_{z}.
\end{split}
\ee
The non-zero NP variables are
\be
\begin{split}
&\Psi_2=-\frac{2M}{(r + M)^3},\\
&\alpha=-\frac{\bz}{2\sqrt2 (r-M)},\quad \xbar\alpha=-\frac{z}{2\sqrt2 (r+M)},\quad \beta=-\xbar\alpha,\quad \xbar\beta=-\alpha,\\ &\gamma=\frac{M}{2(r+M)^2},\quad \xbar\gamma=\frac{M}{2(r+M)^2},\quad \epsilon=\frac{M}{r^2-M^2},\quad \xbar\epsilon=-\frac{M}{r^2-M^2},\\
&\mu=-\frac12 \frac{f}{r+M},\quad \xbar\mu=-\frac12 \frac{f}{r-M},\quad
\rho=-\frac{1}{r+M},\quad \xbar\rho=-\frac{1}{r-M}.
\end{split}
\ee
To have the spin coefficients in the NU gauge, we need the following three rotations of the basis vectors. First, a second class of rotation
\be
b=\frac{2\sqrt2 M (1+\bz+z \bz^2)}{(r+M)\bz},\quad \bar b=0,
\ee
sets $\rho=\xbar\rho$ without turning on $\kappa$ nor $\xbar\kappa$. Then a third class of rotation
\be
A=1,\quad \vartheta=-\ln(1+\bz+z\bz^2),
\ee
turns off $\epsilon$ and $\xbar\epsilon$. The last one is a first class of rotation
\be
a=-\frac{\sqrt2 M}{\bz(r+M)},\quad \bar a=0,
\ee
which turns off $\xbar\pi$ arisen by the previous second class of rotation. The spin coefficients after the rotations become
\begin{align*}
&\Psi_0=-\frac{96M^3}{(r+M)^5\bz^2},\quad \Psi_1=-\frac{12\sqrt2 M^2}{(r+M)^4\bz},\quad \Psi_2=-\frac{2M}{(r+M)^3},\\
&\rho=\xbar\rho=\frac{1}{M-r},\quad \xbar\mu =\frac{1}{2(M-r)},\quad \mu=\frac{M-r}{2(r+M)^2},\\
&\tau=\frac{2\sqrt2 M^2(3r-M)}{(r-M)(r+M)^3\bz},\quad \gamma=\frac{Mr}{(r-M)(r+M)^2},\quad \xbar\gamma=-\frac{Mr}{(r-M)(r+M)^2},\\
&\xbar\nu=\frac{\sqrt2 M^2}{(r-M)(r+M)^2\bz},\quad \alpha=\frac{\bz}{2\sqrt2 (M-r)},\quad \xbar\beta=-\alpha,\\
&\beta=\frac{1}{2\sqrt2 (r+M)^3\bz(1+\bz+z\bz^2)^2)}\bigg[r^2\bz(1+z+4z\bz+3z^2\bz^2)-2Mr(2+3\bz\\
&\hspace{1cm}-z\bz+2\bz^2+4z\bz^3-3z^2 \bz^3+2z^2\bz^4) + M^2(8+17\bz+z\bz+8\bz^2+20z\bz^2\\
&\hspace{1cm}+16z\bz^3+3z^2\bz^3+8z^2\bz^4)\bigg],\quad \xbar\alpha=\tau-\beta,\\
&\xbar\lambda=\frac{M}{(M-r)(r+M)^3\bz^2(1+\bz+z\bz^2)^2}\bigg[4Mr(1+\bz+z\bz^2)^2+r^2(1-2z\bz^2\\
&\hspace{1cm}-2z^2\bz^3)+M^2(2z\bz^2+2z^2\bz^3-1)\bigg],\\
&\sigma=\frac{2M}{(r+M)^4\bz^2(1+\bz+z\bz^2)^2}\bigg[r^2(2z\bz^2+2z^2\bz^3-1)-2Mr(3+4\bz+2\bz^2\\
&\hspace{1cm}+2z\bz^2+4z\bz^3-2z^2\bz^3+2z^2\bz^4)+M^2(7+16\bz+8\bz^2+18z\bz^2+16z\bz^3\\
&\hspace{1cm}+2z^2\bz^3+8z^2\bz^4)\bigg]
\end{align*}
The resulting null basis
\be\label{l}
l=\td \tilde u - \frac{2M \bz}{1+z\bz}\td z + \frac{2M\left[2+(2+z)\bz + 2 z \bz^2\right]}{\bz(1+z\bz)}\td \bz
\ee
should be the gradient of a scalar field. We can define $l=du$. Then $u$ can be solved out by integrating the right hand side of \eqref{l}. The solution is
\be
u=\tilde u + 4 M (\bz+\ln\bz)-2M\ln(1+z\bz).
\ee
In the $(u,r,z,\bz)$ coordinates, the null bases are given by
\be
\begin{split}
&l=\td u,\\
&n= \frac{f}{2}\td u + \td r -\frac{4 M r \left(z \bz^2+\bz+1\right)}{\bz (r+M) (1+ z \bz)} \td \bz,\\
&m=-\frac{2 \sqrt{2} M^2}{\bz (r+M)^2}\td u + \frac{2 \sqrt{2} M}{\bz(r+M)}\td r+\frac{\sqrt{2} (r-M)}{(1+ z \bz)\left(z \bz^2+\bz+1\right)}\td z\\
&\hspace{4cm} +\frac{4 \sqrt{2} M^2 (M-r) \left(z \bz^2+\bz+1\right)}{\bz^2 (r+M)^2 (1+ z \bz)}\td \bz,\\
&\bm=\frac{\sqrt{2} (r+M) \left(z \bz^2+\bz+1\right)}{(1+z\bz)} \td \bz.
\end{split}
\ee
The last operation is to define a new angular variable $\hat z$ to put the tetrad $m$ in the NU gauge. Considering $z$ as function of $(r,\hat z,\bz)$, the only constraint on $z$ is
\be
\frac{\p z}{\p r}= \frac{2 M(1+ z \bz)\left(z \bz^2+\bz+1\right)}{\bz(M^2-r^2)}.
\ee
One possible solution is
\be
z=\frac{1-(1+\bz) e^{2 \arctanh \left(\frac{r}{M}\right)+\frac{\bz \hat{z}}{\sqrt{2}}}}{\bz \left(\bz e^{2 \arctanh \left(\frac{r}{M}\right)+\frac{\bz \hat{z}}{\sqrt{2}}}-1\right)}.
\ee
In the new coordinates $(u,r,\hat z,\bz)$, the basis $m$, now denoted by $\hat m$, becomes
\begin{multline}
\hat m=-\frac{2 \sqrt{2} M^2}{\bz (r+M)^2}\td u +\frac{\sqrt{2} (r-M)}{(1+ z \bz)\left(z \bz^2+\bz+1\right)}\frac{\p z}{\p \hat z}\td \hat z\\
+ \left[\frac{\sqrt{2} (r-M)}{(1+ z \bz)\left(z \bz^2+\bz+1\right)}\frac{\p z}{\p \bz}  +\frac{4 \sqrt{2} M^2 (M-r) \left(z \bz^2+\bz+1\right)}{\bz^2 (r+M)^2 (1+ z \bz)}\right] \td \bz.
\end{multline}
For the rest null bases and the spin coefficients, one just needs to insert $z$ as function in the new coordinates into the expression.


\section{Conclusion}

In this paper, we find that the self-dual condition will not reduce the asymptotic symmetries at null infinity in four dimensions which is still the full BMS symmetry rather than a subset. The reduction of the asymptotic symmetry in \cite{Campiglia:2021srh} is due to the light-cone conditions. The subset BMS symmetry can be recovered from the full BMS symmetry by certain reductions in the self-dual solution space (see also \cite{Mao:2021kxq} for relevant discussions). The self-dual gravity was known to have an infinite ladder of symmetries \cite{Penrose:1968me,Penrose:1976js,Boyer:1985aj}, which can be seen as the origin of recently discovered celestial $w_{1+\infty}$ symmetries \cite{Adamo:2021lrv}. The subset of BMS symmetry in \cite{Campiglia:2021srh} represents two rungs in the infinite ladder. Since the self-dual gravity has the full BMS symmetry at null infinity, one would naturally expect if this indicates another copy of  $w_{1+\infty}$ symmetries. This is a very remarkable question that should be stressed elsewhere.

Asymptotic symmetries are normally studied in certain gauge systems. It is very important for the application of asymptotic symmetries to have exact solutions in the corresponding gauge system. In this paper, we present a generic treatment to transform exact solutions into the NU gauge which has been tested by checking the self-dual Taub-NUT solution. It would be of interest to test our proposal in more generic cases such as the Taub-NUT solution where transformations in series expansion can be found in
\cite{Godazgar:2019dkh,Godazgar:2019ikr} or even the more generic one, the Kerr-Taub-NUT solution. If one is restricted in the self-dual Kerr-Taub-NUT case, a surprising fact is that it can be obtained from the self-dual Taub-NUT solution simply by coordinates transformation \cite{Crawley:2021auj}. It would be very meaningful to study such transformation in the asymptotic framework and to investigate its relation to the BMS transformations.

\section*{Acknowledgments}

The authors thank Glenn Barnich and Jun-Bao Wu for useful discussions. P.M. would like to thank Glenn Barnich again for long term collaborations and supports in relevant research topics. This work is supported in part by the National Natural Science Foundation of China under Grant No. 11905156 and No. 11935009.


\appendix

\section{Self-dual solutions in NP formalism}
\label{solution}

The asymptotic expansions of all NP variables are given by:
\begin{align}
&\Psi_0=\frac{\Psi_0^0(u,z,\bz)}{r^5}+O(r^{-6}),\quad \Psi_1=\frac{\Psi_1^0(u,z,\bz)}{r^4}-\frac{\p_z \Psi_0^0}{r^5}+O(r^{-6}),\nn\\
&\Psi_2=\frac{\Psi_2^0(u,z,\bz)}{r^3}-\frac{\p_z \Psi_1^0}{r^4}+O(r^{-5}),\quad \Psi_3=\frac{\Psi_3^0(u,z,\bz)}{r^2}-\frac{\p_z\Psi_2^0}{r^3}+O(r^{-4}),\nn\\
&\Psi_4=\frac{\Psi_4^0(u,z,\bz)}{r}-\frac{\p_z \Psi_3^0}{r^2}+O(r^{-3}),\nn\\
&\rho=-\frac{1}{r}-\frac{\sigma^0\xbar\sigma^0}{r^3}+O(r^{-5}),\quad \tau=-\frac{\Psi^0_1}{2r^3}+O(r^{-4}),\quad \xbar\tau= O(r^{-4}),\nn\\
&\sigma=\frac{\sigma^0(u,z,\bz)}{r^2}+ \frac{\sigma^0\sigma^0\xbar\sigma^0 - \frac12 \Psi_0^0}{r^4} + O(r^{-5}),\quad \xbar\sigma=\frac{\xbar\sigma^0(u,z,\bz)}{r^2}+ \frac{\sigma^0\xbar\sigma^0\xbar\sigma^0}{r^4} + O(r^{-5}),\nn\\
&\alpha=O(r^{-4}),\quad \xbar\alpha=O(r^{-4}),\nn\\
&\beta=-\frac{\half \Psi^0_1}{r^3}+O(r^{-4}),\quad \xbar\beta=O(r^{-4}),\nn\\
&\mu= - \frac{\sigma^0\lambda^0+\Psi^0_2}{r^2}+ \frac{ \p_z \Psi_1^0}{2r^3} + O(r^{-4}),\quad \xbar\mu= - \frac{\xbar\sigma^0\xbar\lambda^0}{r^2} + O(r^{-4}),\nn\\
&\lambda=\frac{\lambda^0(u,z,\bz)}{r}+ \frac{\sigma^0\xbar\sigma^0 \lambda^0 + \frac12 \xbar\sigma^0 \Psi_2^0}{r^3} + O(r^{-4}),\quad \xbar\lambda=\frac{\xbar\lambda^0(\bz)}{r}+ \frac{\sigma^0\xbar\sigma^0 \lambda^0 }{r^3} + O(r^{-4}),\nn\\
&\gamma=-\frac{\Psi^0_2}{2r^2}+\frac{ \p_z\Psi_1^0}{3r^3} + O(r^{-4}),\quad \xbar\gamma=O(r^{-4}),\nn\\
&\nu=-\frac{\Psi^0_3}{r}+\frac{\p_z \Psi^0_2}{2r^2}+O(r^{-3}),\quad \xbar\nu = O(r^{-3}),\nn\\
&\nn\\
&X^z=\frac{\Psi_1^0}{6r^3}+O(r^{-4}),\quad X^{\bz}=O(r^{-4}), \quad \omega=\frac{\p_z \sigma^0}{r}-\frac{\sigma^0\p_{\bz} \xbar\sigma^0+\half \Psi^0_1}{r^2}+O(r^{-3}),\nn\\
&\xbar\omega=\frac{\p_{\bz} \xbar\sigma^0}{r}-\frac{\xbar\sigma^0\p_z \sigma^0}{r^2}+O(r^{-3}),\quad U=-\frac{\Psi^0_2}{2r}+\frac{\p_z \Psi_1^0}{6r^2} +  O(r^{-3}),\nn\\
&L^z=-\frac{\sigma^0}{r^2} - \frac{\sigma^0\sigma^0\xbar\sigma^0 - \frac16\Psi_0^0}{r^4} +O(r^{-5}),\;\;\;\; L^{\bz}=\frac{1}{r}+\frac{\sigma^0 \xbar\sigma^0 }{r^3}+O(r^{-5}),\nn\\
&\bL^{\bz}=-\frac{\xbar\sigma^0}{r^2} - \frac{\sigma^0\xbar\sigma^0\xbar\sigma^0}{r^4} +O(r^{-5}),\quad \bL^{z}=\frac{1}{r}+\frac{\sigma^0 \xbar\sigma^0 }{r^3}+O(r^{-5}),\nn\\
&L_z=-r+O(r^{-3}),\quad L_{\bz}=-\sigma^0+ \frac{\Psi_0^0}{6r^2} +O(r^{-3}),\nn\\
&\bL_{\bz}=-r+O(r^{-3}),\;\;\;\;\;\; \bL_z=-\xbar\sigma^0  +O(r^{-3}),\nn
\end{align}
where
\begin{align}
&\lambda^0= \p_u \xbar\sigma^0,\quad \sigma^0=u \xbar\lambda^0  + \Theta(z,\bz), \nn\\
&\Psi_2^0= \p_z^2\Theta -\p_{\bz}^2\xbar\sigma^0 + \xbar\sigma^0\xbar\lambda^0 - (u \xbar\lambda^0  + \Theta)\p_u\xbar\sigma^0,\nn\\
&\Psi^0_3=- \p_{\bz}  \p_u \xbar\sigma^0,\quad \Psi^0_4= - \p_u^2  \xbar\sigma^0 ,\nn\\
&\nn\\
&\p_u\Psi^0_0 =\p_{\bz}\Psi^0_1+3(u \xbar\lambda^0  + \Theta)\Psi^0_2,\quad \p_u\Psi^0_1=\p_{\bz}\Psi^0_2+2(u \xbar\lambda^0  + \Theta)\Psi^0_3.\nn
\end{align}


\section{Taub-NUT solutions in NP formalism}
\label{taubnutnp}

Defining
\begin{multline}
\tilde u=t - r - (M+\sqrt{M^2+N^2})\ln (r-M-\sqrt{M^2+N^2})\\
 - (M - \sqrt{M^2+N^2})\ln (r-M+\sqrt{M^2+N^2}),
\end{multline}
we choose the null bases for the metric \eqref{Taub-NUT-metric} as
\be
\begin{split}
&l=\td \tilde u - 2iN \frac{\bz \td z}{1+z\bz} + 2iN \frac{z\td \bz}{1+z\bz},\\
&n= f \left(\frac12\td \tilde u + \frac{\td r}{f} -i N \frac{\bz \td z}{1+z\bz} + iN \frac{z\td \bz}{1+z\bz}\right),\\
&m=\sqrt2(r - i N)\frac{\td z }{(1+z\bz)},\\
&\bm=\sqrt2(r + iN )\frac{\td \bz}{(1+z\bz)}.
\end{split}
\ee
In the vector form, they are
\be
\begin{split}
&l=\p_r,\\
&n= \p_{\tilde u} - \frac{f}{2} \p_r,\\
&m=\frac{i\sqrt2 N z}{r+ iN}\p_{\tilde u} -  \frac{1+z \bz}{\sqrt2 (r+ i N)}\p_{\bz},\\
&\bm=-\frac{i\sqrt2 N \bz}{r- iN}\p_{\tilde u} - \frac{1+z \bz}{\sqrt2 (r - i N)}\p_z.
\end{split}
\ee
The non-zero NP variables are
\be
\begin{split}
&\alpha=-\frac{\bz}{2\sqrt2 (r-iN)},\quad \beta=-\xbar\alpha,\quad \gamma=\frac{r^2M+2N^2 r -M N^2}{2(r+N^2)^2},\\  &\epsilon=\frac{iN}{r^2+N^2},
\quad \mu=-\frac12 \frac{f}{r+iN},\quad
\rho=-\frac{1}{r+iN},\quad \Psi_2=-\frac{M+iN}{(r + iN)^3}.
\end{split}
\ee
The null basis vectors $l$ and $n$ are repeated principal null directions without shear but with twist \cite{Griffiths:2009dfa}.

\section{Three classes of tetrad rotations in Kleinian signature}
\label{rotations}

First class of rotation:
\be
l\to l,\quad m\to m+al,\quad  \bm\to \bm+\ba l,\quad n\to n + a\bm + \ba m + a\ba l.
\ee
One should consider $a$ and $\ba$ as independent real functions. The effect of the transformation on the NP variables are as follows
\be
\begin{split}
&\Psi_0\to\Psi_0,\quad \Psi_1\to \Psi_1+\ba\Psi_0,\quad \Psi_2\to\Psi_2+2\ba\Psi_1+\ba^2\Psi_0,\\
&\Psi_3\to\Psi_3+3\ba\Psi_2+3\ba^2\Psi_1+\ba^3\Psi_0,\\
&\Psi_4\to\Psi_4+4\ba\Psi_3+6\ba^2\Psi_2+4\ba^3\Psi_1+\ba^4\Psi_0,
\end{split}
\ee
and
\begin{equation}
\begin{split}
& \kappa\to \kappa,\quad \sigma\to \sigma+a\kappa,\quad
    \rho\to\rho+\ba\kappa, \quad \epsilon\to \epsilon+\ba\kappa,\\
&\tau\to\tau + a \rho + \ba \sigma + a \ba \kappa,\quad
    \pi\to \pi + 2\ba \epsilon + \ba^2\kappa + l^\mu\p_\mu  \ba,\\
& \alpha\to\alpha + \ba \rho + \ba \epsilon + \ba^2\kappa,\quad \beta\to\beta+a\epsilon + \ba \sigma + a\ba \kappa,\\
& \gamma\to\gamma+a\alpha + \ba\beta + \ba\tau+ a\ba(\rho+\epsilon) + \ba^2\sigma + a \ba^2\kappa,\\
& \lambda\to\lambda+\ba(2\alpha+\pi)+\ba^2(\rho+2\epsilon)+\ba^3\kappa+\bm^\mu\p_\mu\ba + \ba l^\mu\p_\mu \ba,\\
&\mu\to\mu+a\pi+2\ba\beta+2a\ba\epsilon+\ba^2\sigma+a\ba^2\kappa+m^\mu\p_\mu\ba+a l^\mu\p_\mu\ba,\\
& \nu\to\nu+a\lambda+\ba(\mu+2\gamma)+\ba^2(\tau+2\beta)+\ba^3\sigma+a\ba(\pi+2\alpha)\\
&\hspace{1cm} +a\ba^2(\rho+2\epsilon)+a\ba^3\kappa+(n^\mu \p_\mu + \ba m^\mu \p_\mu + a \bm^\mu \p_\mu + a\ba l^\mu \p_\mu)\ba.
\end{split}
\end{equation}
For the other half variables, one just needs to add an overhead bar and swapping $a\rightleftharpoons\ba$.

\noindent Second class of rotation:
\be
n\to n,\quad m\to m+bn,\quad l\to l + b\bar m + \bar b m + b\bb n.
\ee
Here $b$ and $\bb$ are independent real functions. The effect of the transformation on the NP variables are as follows
\be
\begin{split}
&\Psi_4\to\Psi_4,\quad \Psi_3\to \Psi_3+b\Psi_4,\quad \Psi_2\to\Psi_2+2b\Psi_3+b^2\Psi_4,\\
&\Psi_1\to\Psi_1+3b\Psi_2+3b^2\Psi_3+b^3\Psi_4,\\
&\Psi_0\to\Psi_0+4b\Psi_1+6b^2\Psi_2+4b^3\Psi_3+b^4\Psi_4,
\end{split}
\ee
and
\begin{equation}
  \begin{split}
    & \nu\to \nu,\quad \lambda\to \lambda+\bb\nu,\quad
    \mu\to\mu+b\nu, \quad \gamma\to \gamma+b\nu,\\
& \pi\to \pi+b\lambda+
    \bb\mu+b\bb\nu,\quad
\tau\to\tau+2b\gamma+b^2\nu-n^\mu\p_\mu b,\\
&\alpha\to\alpha+b\lambda+\bb\gamma+b\bb\nu,\quad
    \beta\to\beta+b\gamma+b\mu+b^2\nu,\\
& \epsilon\to\epsilon+\bb\beta + b \alpha + b \pi +b\bb(\mu+\gamma) + b^2 \lambda + \bb b^2 \nu,\\
& \sigma\to\sigma + b(2\beta+\tau) + b^2(\mu+2\gamma)+b^3\nu - m^\mu \p_\mu b - b n^\mu \p_\mu b,\\
& \rho\to \rho + \bb \tau + 2 b \alpha + 2 b\bb\gamma+b^2\lambda + \bb b^2 \nu - \bm^\mu\p_\mu b - \bb n^\mu \p_\mu b,\\
& \kappa\to\kappa + \bb \sigma + b(\rho+2\epsilon) + b^2(\pi+2\alpha) + b^3 \lambda + b\bb(\tau+2\beta)\\
&\hspace{1cm}+\bb b^2(\mu+2\gamma)+\bb b^3 \nu - (l^\mu\p_\mu + \bb m^\mu\p_\mu + b\bm^\mu \p_\mu + b \bb n^\mu \p_\mu)b.
\end{split}
\end{equation}
For the other half variables, one just needs to add an overhead bar and swapping $b\rightleftharpoons\bb$.

\noindent Third class of rotation:
\be
l\to \frac{l}{A},\quad n\to A n,\quad m\to e^\vartheta m,\quad  \bm\to e^{-\vartheta}\bm,
\ee
where $A$ and $\vartheta$ are real functions. The effect of the transformation on the NP variables are as follows
\be
\begin{split}
&\Psi_0\to\frac{e^{2\vartheta}}{A^2}\Psi_0,\quad \Psi_1\to \frac{e^{\vartheta}}{A}\Psi_1,\quad \Psi_2\to\Psi_2,\\
&\Psi_3\to\frac{A}{e^{\vartheta}}\Psi_3,\quad \Psi_4\to\frac{A^2}{e^{2\vartheta}}\Psi_4,\\
&\xbar\Psi_0\to\frac{1}{A^2e^{2\vartheta}}\xbar\Psi_0,\quad \xbar\Psi_1\to \frac{1}{A e^{\vartheta}}\xbar\Psi_1,\quad \xbar\Psi_2\to\xbar\Psi_2,\\
&\xbar\Psi_3\to A e^{\vartheta} \xbar\Psi_3,\quad \xbar\Psi_4\to A^2 e^{2\vartheta} \xbar\Psi_4,
\end{split}
\ee
and
\be
\begin{split}
&\kappa\to\frac{e^{\vartheta}}{A^2}\kappa,\quad \xbar\kappa\to\frac{1}{e^{\vartheta}A^2}\xbar\kappa,\quad \sigma\to\frac{e^{2\vartheta}}{A}\sigma,\quad \xbar\sigma\to\frac{1}{A e^{2\vartheta}}\xbar\sigma,\quad \rho\to\frac{1}{A}\rho,\quad \xbar\rho\to\frac{1}{A}\xbar\rho,\\
&\nu =\frac{A^2}{e^{\vartheta}}\nu,\quad \xbar\nu = A^2 e^{\vartheta}\xbar\nu,\quad \lambda=\frac{A}{e^{2\vartheta}}\lambda,\quad \xbar\lambda= A e^{2\vartheta}\xbar\lambda,\quad  \mu=A\mu,\quad \xbar\mu=A\xbar\mu,\\
&\tau=e^{\vartheta}\tau,\quad \xbar\tau=\frac{1}{e^{\vartheta}}\xbar\tau,\quad \pi=\frac{1}{e^{\vartheta}}\pi,\quad \xbar\pi= e^{\vartheta}\xbar\pi,\\
&\gamma\to A\gamma - \frac12 n^\mu\p_\mu A + \frac12 A n^\mu \p_\mu \vartheta,\quad \xbar\gamma\to A\xbar\gamma - \frac12 n^\mu\p_\mu A - \frac12 A n^\mu \p_\mu \vartheta,\\
&\epsilon\to\frac{1}{A}\epsilon + \frac12 l^\mu \p_\mu \frac{1}{A} +  \frac{1}{2A}l^\mu\p_\mu\vartheta,\quad \xbar\epsilon\to\frac{1}{A}\xbar\epsilon + \frac12 l^\mu \p_\mu \frac{1}{A} -  \frac{1}{2A}l^\mu\p_\mu\vartheta,\\
&\alpha\to\frac{1}{e^\vartheta}\alpha - \frac{\bm^\mu}{2 A e^\vartheta}\p_\mu A + \frac{\bm^\mu}{2e^\vartheta} \p_\mu \vartheta,\quad \xbar\alpha\to e^\vartheta\xbar\alpha - \frac{e^\vartheta m^\mu}{2A} \p_\mu A - \frac12  e^\vartheta m^\mu\p_\mu \vartheta,\\
&\beta\to e^\vartheta\beta - \frac{e^\vartheta m^\mu}{2A} \p_\mu A + \frac12  e^\vartheta m^\mu\p_\mu \vartheta, \quad \xbar\beta\to\frac{1}{e^\vartheta}\xbar\beta - \frac{\bm^\mu}{2 A e^\vartheta}\p_\mu A - \frac{\bm^\mu}{2e^\vartheta} \p_\mu \vartheta.
\end{split}
\ee

\bibliography{ref}

\providecommand{\href}[2]{#2}\begingroup\raggedright\begin{thebibliography}{10}

\bibitem{Barnich:2009se}
G.~Barnich and C.~Troessaert, ``{Symmetries of asymptotically flat 4
  dimensional spacetimes at null infinity revisited},''
  \href{http://dx.doi.org/10.1103/PhysRevLett.105.111103}{{\em Phys. Rev.
  Lett.} {\bfseries 105} (2010) 111103},
\href{http://arxiv.org/abs/0909.2617}{{\ttfamily arXiv:0909.2617 [gr-qc]}}.

\bibitem{Barnich:2010eb}
G.~Barnich and C.~Troessaert, ``{Aspects of the BMS/CFT correspondence},''
  \href{http://dx.doi.org/10.1007/JHEP05(2010)062}{{\em JHEP} {\bfseries 05}
  (2010) 062},
\href{http://arxiv.org/abs/1001.1541}{{\ttfamily arXiv:1001.1541 [hep-th]}}.

\bibitem{Barnich:2010ojg}
G.~Barnich and C.~Troessaert, ``{Supertranslations call for superrotations},''
  \href{http://dx.doi.org/10.22323/1.127.0010}{{\em PoS} {\bfseries CNCFG2010}
  (2010) 010}, \href{http://arxiv.org/abs/1102.4632}{{\ttfamily arXiv:1102.4632
  [gr-qc]}}.

\bibitem{Bondi:1962px}
H.~Bondi, M.~G.~J. van~der Burg, and A.~W.~K. Metzner, ``{Gravitational waves
  in general relativity. 7. Waves from axisymmetric isolated systems},''
\href{http://dx.doi.org/10.1098/rspa.1962.0161}{{\em Proc. Roy. Soc. Lond.}
  {\bfseries A269} (1962) 21--52}.

\bibitem{Sachs:1962wk}
R.~K. Sachs, ``{Gravitational waves in general relativity. 8. Waves in
  asymptotically flat space-times},''
\href{http://dx.doi.org/10.1098/rspa.1962.0206}{{\em Proc. Roy. Soc. Lond.}
  {\bfseries A270} (1962) 103--126}.

\bibitem{Strominger:2013jfa}
A.~Strominger, ``{On BMS Invariance of Gravitational Scattering},''
  \href{http://dx.doi.org/10.1007/JHEP07(2014)152}{{\em JHEP} {\bfseries 07}
  (2014) 152}, \href{http://arxiv.org/abs/1312.2229}{{\ttfamily arXiv:1312.2229
  [hep-th]}}.

\bibitem{He:2014laa}
T.~He, V.~Lysov, P.~Mitra, and A.~Strominger, ``{BMS supertranslations and
  Weinberg’s soft graviton theorem},''
  \href{http://dx.doi.org/10.1007/JHEP05(2015)151}{{\em JHEP} {\bfseries 05}
  (2015) 151},
\href{http://arxiv.org/abs/1401.7026}{{\ttfamily arXiv:1401.7026 [hep-th]}}.

\bibitem{Kapec:2014opa}
D.~Kapec, V.~Lysov, S.~Pasterski, and A.~Strominger, ``{Semiclassical Virasoro
  symmetry of the quantum gravity $ \mathcal{S}$-matrix},''
  \href{http://dx.doi.org/10.1007/JHEP08(2014)058}{{\em JHEP} {\bfseries 08}
  (2014) 058}, \href{http://arxiv.org/abs/1406.3312}{{\ttfamily arXiv:1406.3312
  [hep-th]}}.

\bibitem{Campiglia:2014yka}
M.~Campiglia and A.~Laddha, ``{Asymptotic symmetries and subleading soft
  graviton theorem},'' \href{http://dx.doi.org/10.1103/PhysRevD.90.124028}{{\em
  Phys. Rev.} {\bfseries D90} no.~12, (2014) 124028},
\href{http://arxiv.org/abs/1408.2228}{{\ttfamily arXiv:1408.2228 [hep-th]}}.

\bibitem{Guevara:2021abz}
A.~Guevara, E.~Himwich, M.~Pate, and A.~Strominger, ``{Holographic symmetry
  algebras for gauge theory and gravity},''
  \href{http://dx.doi.org/10.1007/JHEP11(2021)152}{{\em JHEP} {\bfseries 11}
  (2021) 152}, \href{http://arxiv.org/abs/2103.03961}{{\ttfamily
  arXiv:2103.03961 [hep-th]}}.

\bibitem{Strominger:2021lvk}
A.~Strominger, ``{w$_{1+\infty}$ and the Celestial Sphere},''
  \href{http://arxiv.org/abs/2105.14346}{{\ttfamily arXiv:2105.14346
  [hep-th]}}.

\bibitem{Ball:2021tmb}
A.~Ball, S.~A. Narayanan, J.~Salzer, and A.~Strominger, ``{Perturbatively exact
  w$_{1+\infty}$ asymptotic symmetry of quantum self-dual gravity},''
  \href{http://dx.doi.org/10.1007/JHEP01(2022)114}{{\em JHEP} {\bfseries 01}
  (2022) 114}, \href{http://arxiv.org/abs/2111.10392}{{\ttfamily
  arXiv:2111.10392 [hep-th]}}.

\bibitem{Ward:1985gz}
R.~S. Ward, ``{Integrable and solvable systems, and relations among them},''
  \href{http://dx.doi.org/10.1098/rsta.1985.0051}{{\em Phil. Trans. Roy. Soc.
  Lond. A} {\bfseries 315} (1985) 451--457}.

\bibitem{Mason:1991rf}
L.~J. Mason and N.~M.~J. Woodhouse, {\em {{Integrability, selfduality, and
  twistor theory}}}.
\newblock Oxford University Press, Oxford, UK, 1996.

\bibitem{Ooguri:1990ww}
H.~Ooguri and C.~Vafa, ``{Selfduality and $N=2$ String {MAGIC}},''
  \href{http://dx.doi.org/10.1142/S021773239000158X}{{\em Mod. Phys. Lett. A}
  {\bfseries 5} (1990) 1389--1398}.

\bibitem{Ooguri:1991fp}
H.~Ooguri and C.~Vafa, ``{Geometry of N=2 strings},''
  \href{http://dx.doi.org/10.1016/0550-3213(91)90270-8}{{\em Nucl. Phys. B}
  {\bfseries 361} (1991) 469--518}.

\bibitem{Kiritsis:1993pb}
E.~Kiritsis, C.~Kounnas, and D.~Lust, ``{A Large class of new gravitational and
  axionic backgrounds for four-dimensional superstrings},''
  \href{http://dx.doi.org/10.1142/S0217751X94000601}{{\em Int. J. Mod. Phys. A}
  {\bfseries 9} (1994) 1361--1394},
  \href{http://arxiv.org/abs/hep-th/9308124}{{\ttfamily arXiv:hep-th/9308124}}.

\bibitem{Bianchi:1994gi}
M.~Bianchi, F.~Fucito, G.~C. Rossi, and M.~Martellini, ``{ALE instantons in
  string effective theory},''
  \href{http://dx.doi.org/10.1016/0550-3213(94)00552-P}{{\em Nucl. Phys. B}
  {\bfseries 440} (1995) 129--170},
  \href{http://arxiv.org/abs/hep-th/9409037}{{\ttfamily arXiv:hep-th/9409037}}.

\bibitem{Berkovits:1994vy}
N.~Berkovits and C.~Vafa, ``{N=4 topological strings},''
  \href{http://dx.doi.org/10.1016/0550-3213(94)00419-F}{{\em Nucl. Phys. B}
  {\bfseries 433} (1995) 123--180},
  \href{http://arxiv.org/abs/hep-th/9407190}{{\ttfamily arXiv:hep-th/9407190}}.

\bibitem{Berkovits:1994ym}
N.~Berkovits, ``{Vanishing theorems for the selfdual N=2 string},''
  \href{http://dx.doi.org/10.1016/0370-2693(95)00308-8}{{\em Phys. Lett. B}
  {\bfseries 350} (1995) 28--32},
  \href{http://arxiv.org/abs/hep-th/9412179}{{\ttfamily arXiv:hep-th/9412179}}.

\bibitem{Duff:1994an}
M.~J. Duff, R.~R. Khuri, and J.~X. Lu, ``{String solitons},''
  \href{http://dx.doi.org/10.1016/0370-1573(95)00002-X}{{\em Phys. Rept.}
  {\bfseries 259} (1995) 213--326},
  \href{http://arxiv.org/abs/hep-th/9412184}{{\ttfamily arXiv:hep-th/9412184}}.

\bibitem{Berkovits:1995ab}
N.~Berkovits, ``{SuperPoincare invariant superstring field theory},''
  \href{http://dx.doi.org/10.1016/0550-3213(95)00259-U}{{\em Nucl. Phys. B}
  {\bfseries 450} (1995) 90--102},
  \href{http://arxiv.org/abs/hep-th/9503099}{{\ttfamily arXiv:hep-th/9503099}}.
  [Erratum: Nucl.Phys.B 459, 439--451 (1996)].

\bibitem{Ooguri:1995cp}
H.~Ooguri and C.~Vafa, ``{All loop N=2 string amplitudes},''
  \href{http://dx.doi.org/10.1016/0550-3213(95)00365-Y}{{\em Nucl. Phys. B}
  {\bfseries 451} (1995) 121--161},
  \href{http://arxiv.org/abs/hep-th/9505183}{{\ttfamily arXiv:hep-th/9505183}}.

\bibitem{Bern:1998xc}
Z.~Bern, L.~J. Dixon, M.~Perelstein, and J.~S. Rozowsky, ``{One loop n point
  helicity amplitudes in (selfdual) gravity},''
  \href{http://dx.doi.org/10.1016/S0370-2693(98)01397-5}{{\em Phys. Lett. B}
  {\bfseries 444} (1998) 273--283},
  \href{http://arxiv.org/abs/hep-th/9809160}{{\ttfamily arXiv:hep-th/9809160}}.

\bibitem{Monteiro:2011pc}
R.~Monteiro and D.~O'Connell, ``{The Kinematic Algebra From the Self-Dual
  Sector},'' \href{http://dx.doi.org/10.1007/JHEP07(2011)007}{{\em JHEP}
  {\bfseries 07} (2011) 007}, \href{http://arxiv.org/abs/1105.2565}{{\ttfamily
  arXiv:1105.2565 [hep-th]}}.

\bibitem{Monteiro:2014cda}
R.~Monteiro, D.~O'Connell, and C.~D. White, ``{Black holes and the double
  copy},'' \href{http://dx.doi.org/10.1007/JHEP12(2014)056}{{\em JHEP}
  {\bfseries 12} (2014) 056}, \href{http://arxiv.org/abs/1410.0239}{{\ttfamily
  arXiv:1410.0239 [hep-th]}}.

\bibitem{Krasnov:2016emc}
K.~Krasnov, ``{Self-Dual Gravity},''
  \href{http://dx.doi.org/10.1088/1361-6382/aa65e5}{{\em Class. Quant. Grav.}
  {\bfseries 34} no.~9, (2017) 095001},
  \href{http://arxiv.org/abs/1610.01457}{{\ttfamily arXiv:1610.01457
  [hep-th]}}.

\bibitem{Berman:2018hwd}
D.~S. Berman, E.~Chac\'on, A.~Luna, and C.~D. White, ``{The self-dual classical
  double copy, and the Eguchi-Hanson instanton},''
  \href{http://dx.doi.org/10.1007/JHEP01(2019)107}{{\em JHEP} {\bfseries 01}
  (2019) 107}, \href{http://arxiv.org/abs/1809.04063}{{\ttfamily
  arXiv:1809.04063 [hep-th]}}.

\bibitem{Campiglia:2021srh}
M.~Campiglia and S.~Nagy, ``{A double copy for asymptotic symmetries in the
  self-dual sector},'' \href{http://dx.doi.org/10.1007/JHEP03(2021)262}{{\em
  JHEP} {\bfseries 03} (2021) 262},
  \href{http://arxiv.org/abs/2102.01680}{{\ttfamily arXiv:2102.01680
  [hep-th]}}.

\bibitem{Newman:1962cia}
E.~T. Newman and T.~W.~J. Unti, ``{Behavior of Asymptotically Flat Empty
  Spaces},''
\href{http://dx.doi.org/10.1063/1.1724303}{{\em J. Math. Phys.} {\bfseries 3}
  no.~5, (1962) 891}.

\bibitem{Brown:1986nw}
J.~D. Brown and M.~Henneaux, ``{Central Charges in the Canonical Realization of
  Asymptotic Symmetries: An Example from Three-Dimensional Gravity},''
  \href{http://dx.doi.org/10.1007/BF01211590}{{\em Commun. Math. Phys.}
  {\bfseries 104} (1986) 207--226}.

\bibitem{Strominger:1997eq}
A.~Strominger, ``{Black hole entropy from near horizon microstates},''
  \href{http://dx.doi.org/10.1088/1126-6708/1998/02/009}{{\em JHEP} {\bfseries
  02} (1998) 009}, \href{http://arxiv.org/abs/hep-th/9712251}{{\ttfamily
  arXiv:hep-th/9712251}}.

\bibitem{Guica:2008mu}
M.~Guica, T.~Hartman, W.~Song, and A.~Strominger, ``{The Kerr/CFT
  Correspondence},'' \href{http://dx.doi.org/10.1103/PhysRevD.80.124008}{{\em
  Phys. Rev. D} {\bfseries 80} (2009) 124008},
  \href{http://arxiv.org/abs/0809.4266}{{\ttfamily arXiv:0809.4266 [hep-th]}}.

\bibitem{Taub:1950ez}
A.~H. Taub, ``{Empty space-times admitting a three parameter group of
  motions},'' \href{http://dx.doi.org/10.2307/1969567}{{\em Annals Math.}
  {\bfseries 53} (1951) 472--490}.

\bibitem{Newman:1963yy}
E.~Newman, L.~Tamburino, and T.~Unti, ``{Empty space generalization of the
  Schwarzschild metric},'' \href{http://dx.doi.org/10.1063/1.1704018}{{\em J.
  Math. Phys.} {\bfseries 4} (1963) 915}.

\bibitem{Newman:1961qr}
E.~Newman and R.~Penrose, ``{An Approach to gravitational radiation by a method
  of spin coefficients},''
\href{http://dx.doi.org/10.1063/1.1724257}{{\em J. Math. Phys.} {\bfseries 3}
  (1962) 566--578}.

\bibitem{Chandrasekhar}
S.~Chandrasekhar, ``{The Newman-Penrose formalism},'' in {\em {The mathematical
  theory of black holes}}, ch.~1, pp.~40--55.
\newblock Oxford, UK, 1983.

\bibitem{Elvang:2013cua}
H.~Elvang and Y.-t. Huang, ``{Scattering Amplitudes},''
  \href{http://arxiv.org/abs/1308.1697}{{\ttfamily arXiv:1308.1697 [hep-th]}}.

\bibitem{Henn:2014yza}
J.~M. Henn and J.~C. Plefka,
  \href{http://dx.doi.org/10.1007/978-3-642-54022-6}{{\em {Scattering
  Amplitudes in Gauge Theories}}}, vol.~883.
\newblock Springer, Berlin, 2014.

\bibitem{Atanasov:2021oyu}
A.~Atanasov, A.~Ball, W.~Melton, A.-M. Raclariu, and A.~Strominger, ``{(2, 2)
  Scattering and the celestial torus},''
  \href{http://dx.doi.org/10.1007/JHEP07(2021)083}{{\em JHEP} {\bfseries 07}
  (2021) 083}, \href{http://arxiv.org/abs/2101.09591}{{\ttfamily
  arXiv:2101.09591 [hep-th]}}.

\bibitem{Crawley:2021auj}
E.~Crawley, A.~Guevara, N.~Miller, and A.~Strominger, ``{Black Holes in Klein
  Space},'' \href{http://arxiv.org/abs/2112.03954}{{\ttfamily arXiv:2112.03954
  [hep-th]}}.

\bibitem{Barnich:2016lyg}
G.~Barnich and C.~Troessaert, ``{Finite BMS transformations},''
  \href{http://dx.doi.org/10.1007/JHEP03(2016)167}{{\em JHEP} {\bfseries 03}
  (2016) 167},
\href{http://arxiv.org/abs/1601.04090}{{\ttfamily arXiv:1601.04090 [gr-qc]}}.

\bibitem{Compere:2016jwb}
G.~Comp\`{e}re and J.~Long, ``{Vacua of the gravitational field},''
  \href{http://dx.doi.org/10.1007/JHEP07(2016)137}{{\em JHEP} {\bfseries 07}
  (2016) 137}, \href{http://arxiv.org/abs/1601.04958}{{\ttfamily
  arXiv:1601.04958 [hep-th]}}.

\bibitem{Ball:2018prg}
A.~Ball, M.~Pate, A.-M. Raclariu, A.~Strominger, and R.~Venugopalan,
  ``{Measuring color memory in a color glass condensate at
  electron\textendash{}ion colliders},''
  \href{http://dx.doi.org/10.1016/j.aop.2019.04.010}{{\em Annals Phys.}
  {\bfseries 407} (2019) 15--28},
  \href{http://arxiv.org/abs/1805.12224}{{\ttfamily arXiv:1805.12224
  [hep-ph]}}.

\bibitem{Barnich:2021dta}
G.~Barnich and R.~Ruzziconi, ``{Coadjoint representation of the BMS group on
  celestial Riemann surfaces},''
  \href{http://dx.doi.org/10.1007/JHEP06(2021)079}{{\em JHEP} {\bfseries 06}
  (2021) 079}, \href{http://arxiv.org/abs/2103.11253}{{\ttfamily
  arXiv:2103.11253 [gr-qc]}}.

\bibitem{Barnich:2011ty}
G.~Barnich and P.-H. Lambert, ``{A note on the Newman-Unti group},'' {\em Adv.
  Math. Phys.} {\bfseries 2012} (2012) 197385,
\href{http://arxiv.org/abs/1102.0589}{{\ttfamily arXiv:1102.0589 [gr-qc]}}.

\bibitem{Compere:2018ylh}
G.~Comp\`ere, A.~Fiorucci, and R.~Ruzziconi, ``{Superboost transitions,
  refraction memory and super-Lorentz charge algebra},''
  \href{http://dx.doi.org/10.1007/JHEP11(2018)200}{{\em JHEP} {\bfseries 11}
  (2018) 200}, \href{http://arxiv.org/abs/1810.00377}{{\ttfamily
  arXiv:1810.00377 [hep-th]}}. [Erratum: JHEP 04, 172 (2020)].

\bibitem{Campiglia:2016efb}
M.~Campiglia and A.~Laddha, ``{Sub-subleading soft gravitons and large
  diffeomorphisms},'' \href{http://dx.doi.org/10.1007/JHEP01(2017)036}{{\em
  JHEP} {\bfseries 01} (2017) 036},
  \href{http://arxiv.org/abs/1608.00685}{{\ttfamily arXiv:1608.00685 [gr-qc]}}.

\bibitem{Freidel:2021dfs}
L.~Freidel, D.~Pranzetti, and A.-M. Raclariu, ``{Sub-subleading soft graviton
  theorem from asymptotic Einstein\textquoteright{}s equations},''
  \href{http://dx.doi.org/10.1007/JHEP05(2022)186}{{\em JHEP} {\bfseries 05}
  (2022) 186}, \href{http://arxiv.org/abs/2111.15607}{{\ttfamily
  arXiv:2111.15607 [hep-th]}}.

\bibitem{Pasterski:2016qvg}
S.~Pasterski, S.-H. Shao, and A.~Strominger, ``{Flat Space Amplitudes and
  Conformal Symmetry of the Celestial Sphere},''
  \href{http://dx.doi.org/10.1103/PhysRevD.96.065026}{{\em Phys. Rev. D}
  {\bfseries 96} no.~6, (2017) 065026},
  \href{http://arxiv.org/abs/1701.00049}{{\ttfamily arXiv:1701.00049
  [hep-th]}}.

\bibitem{Pasterski:2017kqt}
S.~Pasterski and S.-H. Shao, ``{Conformal basis for flat space amplitudes},''
  \href{http://dx.doi.org/10.1103/PhysRevD.96.065022}{{\em Phys. Rev. D}
  {\bfseries 96} no.~6, (2017) 065022},
  \href{http://arxiv.org/abs/1705.01027}{{\ttfamily arXiv:1705.01027
  [hep-th]}}.

\bibitem{Pasterski:2021rjz}
S.~Pasterski, ``{Lectures on celestial amplitudes},''
  \href{http://dx.doi.org/10.1140/epjc/s10052-021-09846-7}{{\em Eur. Phys. J.
  C} {\bfseries 81} no.~12, (2021) 1062},
  \href{http://arxiv.org/abs/2108.04801}{{\ttfamily arXiv:2108.04801
  [hep-th]}}.

\bibitem{Ball:2019atb}
A.~Ball, E.~Himwich, S.~A. Narayanan, S.~Pasterski, and A.~Strominger,
  ``{Uplifting AdS$_{3}$/CFT$_{2}$ to flat space holography},''
  \href{http://dx.doi.org/10.1007/JHEP08(2019)168}{{\em JHEP} {\bfseries 08}
  (2019) 168}, \href{http://arxiv.org/abs/1905.09809}{{\ttfamily
  arXiv:1905.09809 [hep-th]}}.

\bibitem{Pasterski:2021raf}
S.~Pasterski, M.~Pate, and A.-M. Raclariu, ``{Celestial Holography},'' in {\em
  {2022 Snowmass Summer Study}}.
\newblock 11, 2021.
\newblock \href{http://arxiv.org/abs/2111.11392}{{\ttfamily arXiv:2111.11392
  [hep-th]}}.

\bibitem{Barnich:2019vzx}
G.~Barnich, P.~Mao, and R.~Ruzziconi, ``{BMS current algebra in the context of
  the Newman-Penrose formalism},''
  \href{http://dx.doi.org/10.1088/1361-6382/ab7c01}{{\em Class. Quant. Grav.}
  {\bfseries 37} no.~9, (2020) 095010},
  \href{http://arxiv.org/abs/1910.14588}{{\ttfamily arXiv:1910.14588 [gr-qc]}}.

\bibitem{2003CQGra..20.4153F}
S.~J. {Fletcher} and A.~W.~C. {Lun}, ``{The Kerr spacetime in generalized Bondi
  Sachs coordinates},''
  \href{http://dx.doi.org/10.1088/0264-9381/20/19/302}{{\em Classical and
  Quantum Gravity} {\bfseries 20} no.~19, (2003) 4153--4167}.

\bibitem{Griffiths:2009dfa}
J.~B. Griffiths and J.~Podolsky,
  \href{http://dx.doi.org/10.1017/CBO9780511635397}{{\em {Exact Space-Times in
  Einstein's General Relativity}}}.
\newblock Cambridge University Press, Cambridge, 2009.

\bibitem{Mao:2021kxq}
P.~Mao and W.~Zhao, ``{Note on the asymptotic structure of Kerr-Schild form},''
  \href{http://dx.doi.org/10.1007/JHEP01(2022)030}{{\em JHEP} {\bfseries 01}
  (2022) 030}, \href{http://arxiv.org/abs/2109.09676}{{\ttfamily
  arXiv:2109.09676 [gr-qc]}}.

\bibitem{Penrose:1968me}
R.~Penrose, ``{Twistor quantization and curved space-time},''
  \href{http://dx.doi.org/10.1007/BF00668831}{{\em Int. J. Theor. Phys.}
  {\bfseries 1} (1968) 61--99}.

\bibitem{Penrose:1976js}
R.~Penrose, ``{Nonlinear Gravitons and Curved Twistor Theory},''
  \href{http://dx.doi.org/10.1007/BF00762011}{{\em Gen. Rel. Grav.} {\bfseries
  7} (1976) 31--52}.

\bibitem{Boyer:1985aj}
C.~P. Boyer and J.~F. Plebanski, ``{An infinite hierarchy of conservation laws
  and nonlinear superposition principles for selfdual Einstein spaces},''
  \href{http://dx.doi.org/10.1063/1.526652}{{\em J. Math. Phys.} {\bfseries 26}
  (1985) 229--234}.

\bibitem{Adamo:2021lrv}
T.~Adamo, L.~Mason, and A.~Sharma, ``{Celestial $w_{1+\infty}$ Symmetries from
  Twistor Space},'' \href{http://dx.doi.org/10.3842/SIGMA.2022.016}{{\em SIGMA}
  {\bfseries 18} (2022) 016}, \href{http://arxiv.org/abs/2110.06066}{{\ttfamily
  arXiv:2110.06066 [hep-th]}}.

\bibitem{Godazgar:2019dkh}
H.~Godazgar, M.~Godazgar, and C.~N. Pope, ``{Dual gravitational charges and
  soft theorems},'' \href{http://dx.doi.org/10.1007/JHEP10(2019)123}{{\em JHEP}
  {\bfseries 10} (2019) 123}, \href{http://arxiv.org/abs/1908.01164}{{\ttfamily
  arXiv:1908.01164 [hep-th]}}.

\bibitem{Godazgar:2019ikr}
H.~Godazgar, M.~Godazgar, and C.~N. Pope, ``{Taub-NUT from the Dirac
  monopole},'' \href{http://dx.doi.org/10.1016/j.physletb.2019.134938}{{\em
  Phys. Lett. B} {\bfseries 798} (2019) 134938},
  \href{http://arxiv.org/abs/1908.05962}{{\ttfamily arXiv:1908.05962
  [hep-th]}}.

\end{thebibliography}\endgroup

\end{document}